\documentclass[11pt]{article}
\usepackage{amssymb}
\usepackage{epsfig}
\begin{document}
\title{{\bf{\Large Conformal Transformation, Near Horizon Symmetry, Virasoro Algebra and Entropy}}}
\author{ 
 {\bf {\normalsize Bibhas Ranjan Majhi}$
$\thanks{E-mail: bibhas.majhi@mail.huji.ac.il}}\\ 
{\normalsize Racah Institute of Physics, Hebrew University of Jerusalem,}
\\{\normalsize Givat Ram, Jerusalem 91904, Israel}
\\[0.3cm]
}

\maketitle

\begin{abstract}
There are certain black hole solutions in general relativity (GR) which are conformally related to the stationary solutions in GR. It is not obvious that the horizon entropy of these spacetimes is also one quarter of the area of horizon, like the stationary ones. Here I study this topic in the context of Virasoro algebra and Cardy formula. Using the fact that the conformal spacetime admits conformal Killing vector and the horizon is determined by the vanishing of the norm of it, the diffemorphisms are obtained which keep the near horizon structure invariant. The Noether charge and a bracket among them corresponding to these vectors are calculated in this region. Finally, they are evaluated for the Sultana-Dyer (SD) black hole, which is conformal to the Schwarzschild metric. It is found that the bracket is identical to the usual Virasoro algebra with the central extension. Identifying the zero mode eigenvalue and the central charge, the entropy of the SD horizon is obtained by using Cardy formula. Interestingly, this is again one quarter of the horizon area. Only difference in this case is that the area is modified by the conformal factor compared to that of the stationary one. The analysis gives a direct proof of the earlier assumption.  
\end{abstract}

\section{\label{Intro}Introduction}
   In the attempt of quantization of the gravitational theory, it has been arisen that the black holes radiate \cite{Hawking:1974rv}. The radiation is completely quantum mechanical and predicts that the horizon has a temperature which is proportional to the surface gravity.  Correspondingly, one can associate entropy on the horizon, suggested first by Bekenstein through a thought experiment \cite{Bekenstein:1973ur}. In total the black hole solutions behave like a thermodynamic system and the laws are similar to the usual laws of thermodynamics \cite{Bardeen:1973gs}. In the case of general theory of relativity (GR) the entropy is given by the Bekenstein-Hawking area law $S=A/4G$. Till now it has been a central attention of the people to understand the origin of the entropy, more specifically to give a microscopic description, in a hope that it might shed some light towards the quantum description of gravity. Several methods has been proposed so far. But none of the them are complete or self consistent.

   Among the several methods, one important approach is given by Wald \cite{Wald:1993nt,Iyer:1995kg} in which it has been shown that the Noether charge of the gravity theory corresponding to the Killing symmetry, calculated on the horizon, is related to the entropy. It implies that the Killing symmetries have something to do with the degrees of freedom (DOF) responsible for the entropy. Carlip \cite{Carlip:1999cy} made an attempt which is based on the usual Virasoro algebra \cite{Book} and Cardy formula \cite{Cardy:1986ie,Carlip:1998qw}. This was initiated from the original approach of Brown and Henneaux \cite{Brown:1986nw} and Strominger et.al \cite{Strominger:1997eq}. Carlip used the diffeomorphism vectors which keep the Killing horizon structure invariant. It was showed that the Fourier mode of a bracket among the charges is identical to the usual Virasoro algebra. Then identifying the zero mode eigenvalue and the central charge and substituting them in Cardy formula one obtains the entropy of the horizon. Later on people has adopted this approach and investigated its different aspects \cite{Majhi:2012st} (For recent developments, see \cite{Majhi:2011ws,Majhi:2012tf,Majhi:2013lba}). Almost all cases, so far I am aware of, were for the static or stationary spacetimes.

 The most of the earlier calculations adopt one or more unclear assumptions. (For a detailed mention of these, see Introduction of \cite{Majhi:2011ws}). In one of my earlier works with T. Padmanabhan \cite{Majhi:2011ws}, we gave a definition of bracket among the charges which is free of these ambiguities. It is based only on the fact that {\it the Noether current $J^a$ of any diffeomorphism invariant theory can be expressed as the covariant derivative of the Noether potential $J^{ab}$}. It has been shown \cite{Majhi:2011ws,Majhi:2012tf} that such a definition yields Virasoro like algebra with the central extension for the ``horizon structure invariant diffeomorphisms'' which gives rise to the usual form of entropy by Cardy formula. The significance of such result is as follows. Some of the DOFs, which were originally gauge DOF, raised to physical DOF which lead to the entropy. This is happening due to the choice of the particular class of diffeomorphism vectors. Moreover, the entropy is now observer dependent quantity and the DOFs, which are responsible for it, have no absolute notion. This has been elaborated in \cite{Majhi:2012tf,Majhi:2012nq,Majhi:2013jpk}.

In this paper, I shall evaluate the entropy of spacetimes which are conformal to the stationary black hole solutions of the gravity theory in the context of Virasoro algebra and Cardy formula. The discussion will be confined within GR and we will consider the conformal spacetime which is a solution of GR itself. The analysis will be inspired by the approach of Carlip \cite{Carlip:1999cy}. Since the conformal metric is also a solution of the GR theory, the Noether current and the charge are well known. Also the definition of the bracket among the charges will be taken from our analysis \cite{Majhi:2011ws}. Next it is necessary to find the relevant diffeomorphism vectors to evaluate them explicitly. This will be chosen as follows. 
Since the seed metric is stationary, it has a timelike Killing vector. Now it can be shown that the same vector acts as the conformal Killing vector for the conformally related metric \cite{Jacobson:1993pf}. The horizon is defined by the vanishing of the norm of the vector. Therefore, we will look for those diffemorphisms which will keep conformal Killing horizon structure invariant after the perturbation. This will be cleared when we go into the main discussion.

  After having the diffeomorphisms, we shall calculate the charge and the bracket in the near horizon limit. Through out the analysis, our aim will be to find the relation between the above quantities with those for the usual stationary case. This will give not only a mapping of the entities under conformal transformation, it will also simplify the evaluation extensively since the stationary calculation is already known from earlier literature \cite{Carlip:1999cy,Majhi:2011ws}. Having all the necessary quantities the central term will be found out. Then we shall calculate the Fourier modes. Till now the analysis will be general. Next the explicit expressions for the Fourier modes of the charge and the central term will be given for the Sultana-Dyer (SD) black hole. We shall show that these satisfy an algebra which is similar to the usual Virasoro form. Identifying the zero mode eigenvalue and the central charge and then substituting in the Cardy formula we shall obtain the entropy of the SD horizon. We will see that the entropy is again one quarter of the horizon area; i.e. it obeys the area law. The only difference from the stationary one is that the area is modified by the conformal factor.

  Let me now state why such an analysis is necessary and what are the implications of it. These are as follows:\\
$\bullet$ There exists solutions which are related to the usual popular stationary solutions by a conformal factor. Till now an extensive studies have been done to explore the thermodynamics of these stationary solutions and so their thermodynamic structure is well known. Therefore lack of discussions for the conformal metric can be filled by just relating the relevant quantities under the conformal transformation. This will make life simple.\\
$\bullet$ The conformal metric, in the present paper, will be considered as the SD spacetime. This is a black hole solution of GR in presence of two noninteracting fluids. One is timelike and the other is null-like \cite{Sultana:2005tp}. Moreover, it is related to the Schwarzschild spacetime by a time dependent conformal factor and so the metric is an evolving one. Also the asymptotic structure is Friedmann-Lemaitre-Robertson-Walker (FLRW). Therefore the usual stationary prescription or the asymptotic flat prescription can not be applicable to revel the thermodynamics without proper modifications. But for the realistic purpose it is necessary to explore the time dependent cases, since our universe is evolving. The present analysis, so far I am aware of, will give the first application of the Virasoro algebra and Cardy formula methodology to find entropy for a time dependent spacetime.\\
$\bullet$ In literature, the entropy of the SD metric has been taken to be determined by the area law without any explicit derivation (For example, see \cite{Faraoni:2007gq,Majhi:2014hpa}). Intuitively this may be fine since SD is also a solution of GR and in GR the entropy is given by Bekenstein-Hawking area law. But most of the analysis were for stationary background. Therefore in absence of any direct proof, such stationary result can not be borrowed for evolving one without any concrete justification. This analysis will demonstrate a calculation of entropy for the SD spacetime and fill this gap.\\
From the above discussion, it is obvious that the present paper will address some relevant issues in the present context.

  The plan of the paper is as follows: In section \ref{generators}, I will introduce the condition to maintain the conformal horizon structure from which the relevant diffeomorphism vectors will be determined. Using these vectors the Noether charge and the bracket will be evaluated near the horizon in the next section. We shall also find the central term in the same section. Section \ref{Fourier} will present the Fourier modes of the charge and the central term. Then in next section they will be calculated for the SD metric and also shown to give the Virasoro algebra. Identifying the relevant quantities the entropy of SD spacetime will be obtained in this section. Final section will be devoted for the conclusions and discussions. I shall also give three appendices at the end of the paper which will contain details of the main calculation. 
We shall adopt the following notations and symbols in the paper. The Latin indices $a,b,c,$ etc. denote all the spacetime indices. The unbar quantities are for the seed metric while the bar ones correspond to the conformally connected metric. 

\section{\label{generators}Setup: conformal transformation and generators for the near horizon symmetry}
   In this section, the relevant diffeomorphism vectors will be chosen such that the conformal Killing horizon structure remains invariant. The discussion will be restricted to those spacetimes whose seed matrices are stationary so that the Killing vectors of the seed matrices turn out to be the conformal Killing vector for them. In addition the horizon is defined by the vanishing of the norm of the vector. Also we shall find the conditions for the closer algebra, satisfied by the diffeomorphism vectors. These are necessary for the subsequent analysis. The whole discussion will be in the line of the earlier analysis \cite{Carlip:1999cy,Majhi:2011ws} for the stationary spacetimes which admit Killing horizon. Here the difference is that the spacetime has conformal Killing vector instead of timelike Killing one.

  Consider a stationary seed metric $g_{ab}$. Then it must has a timelike Killing vector $\chi^a$ so that $\pounds_{\chi}g_{ab}=0$. Now choose our metric $\bar{g}_{ab}$ such that it is related to the earlier one by a conformal factor; i.e. $\bar{g}_{ab}=\Omega^2g_{ab}$. In this case it has been shown in \cite{Jacobson:1993pf} that $\bar{\chi}^a=\chi^a$ is the conformal Killing vector for $\bar{g}_{ab}$; i.e. $\bar{\chi}^a$ satisfies 
\begin{equation}
\pounds_{\bar{\chi}}\bar{g}_{ab} = (\pounds_{\bar{\chi}}\Omega^2) g_{ab} = (\pounds_{\bar{\chi}}\ln\Omega^2)
\bar{g}_{ab}~,
\label{2.01} 
\end{equation}
where $\pounds_{\bar{\chi}}$ is the Lie derivative along $\bar{\chi}^a$.
Also note that $\bar{\chi}_a = \bar{g}_{ab}\bar{\chi}^b=\Omega^2g_{ab}\chi^b=\Omega^2\chi_a$. {\it Remember that the unbar tensors are raised or lowered by $g_{ab}$ whereas raising or lowering of the bar ones are done by $\bar{g}_{ab}$}. 
The initial seed metric has a Killing horizon which is determined by $g_{ab}\chi^a\chi^b\equiv\chi^2=0$. Interestingly, the same remains as the horizon for $\bar{g}_{ab}$ since $\bar{g}_{ab}\bar{\chi}^a\bar{\chi}^b=\Omega^2\chi^2 =0$ when $\chi^2=0$, provided $\Omega^2\neq\mathcal{O}(\chi^{-2})$. This we shall call as {\it conformal killing horizon}. Then one can define surface gravities for the two matrices by the following relations:
\begin{equation}
\nabla_a(\chi^2) = -2\kappa\rho_a; \,\,\ \bar{\nabla}_a(\bar{\chi}^2) = -2\bar{\kappa}\bar{\rho}_a
\label{2.02} 
\end{equation}
where $\rho^a$ and $\bar{\rho}^a=\rho^a$ are the normal vectors on the Killing and conformal Killing horizons, respectively. In the near horizon limit one has $\rho^a\rightarrow\chi^a$ and $\bar{\rho}^a\rightarrow\bar{\chi}^a$. Here, we denote $\bar{\chi}^2$ as $\bar{\chi}^2 \equiv \bar{g}_{ab}\bar{\chi}^a\bar{\chi}^b$ and $\bar{\rho}_a = \Omega^2\rho_a$. It can be checked that $\rho_a\chi^a=0$ and $\bar{\rho}_a\bar{\chi}^a=0$. In principle we can have a relation between $\kappa$ and $\bar{\kappa}$. This is found out to be $\bar{\kappa}=\kappa - (\chi^2/\rho^2)\rho^a\nabla_a\ln\Omega^2$.

   Let us now introduce the diffeomorphism generators $\bar{\xi}^a$ which leave the asymptotic conformal Killing horizon structure invariant after the perturbation near the horizon. This will be chosen by imposing the following condition:
\begin{equation}
\frac{\bar{\chi}^a\bar{\chi}^b\pounds_{\bar{\xi}}\bar{g}_{ab} - \bar{\chi}^a\bar{\chi}^b(\pounds_{\bar{\xi}}
\ln\Omega^2)\bar{g}_{ab}}{\bar{\chi}^2} \rightarrow 0
\label{2.03}
\end{equation}
near the horizon. Using the relations of the relevant quantities between the two matrices, it is possible to cast the above as
\begin{equation}
\frac{\chi^a\chi^b\pounds_{\bar{\xi}}g_{ab}}{\chi^2} \rightarrow 0~.
\label{2.04} 
\end{equation}
Interestingly, the similar condition in form was also invoked by Carlip for the stationary case \cite{Carlip:1999cy}. But we shall see later that diffeomorphism vector will turn out to be different in this case due to the presence of the conformal factor.
As earlier, consider $\bar{\xi}^a$ such that it talks about only the diffeomorphism of the ($t-r$) sector of the metric. Therefore take the form of it as
\begin{equation}
\bar{\xi^a} = T\bar{\chi}^a + R\bar{\rho}^a
\label{2.05} 
\end{equation}
where $T$ and $R$ are two unknown functions. Note that $\bar{\xi}^a = T \chi^a+R\rho^a\equiv\xi^a$ whereas $\bar{\xi}_a = T\bar{\chi}_a+R\bar{\rho}_a = \Omega^2\xi_a$. Using this in (\ref{2.04}) we obtain a relation between the two unknown functions:
\begin{equation}
R=\frac{\chi^2}{\kappa\rho^2}DT +  \frac{\chi^2}{\kappa\rho^2}TD(\ln \Omega^2)~,
\label{2.06}
\end{equation}
with $D\equiv\chi^a\nabla_a$. Observe that the first term is same as earlier \cite{Carlip:1999cy,Majhi:2011ws} while the last term appears due to the conformal transformation. Now $\bar{\xi}^a$, given by (\ref{2.05}) and (\ref{2.06}), will preserve the closer relation 
\begin{equation}
\{\bar{\xi}_1,\bar{\xi}_2\}^a = \{T_1,T_2\}\bar{\chi}^a + \{R_1,R_2\}\bar{\rho}^a 
\label{2.07}
\end{equation}
near the horizon when the following conditions are satisfied:
\begin{equation}
\rho^a\nabla_a T =0; \,\,\,\ \rho^a\nabla_aDT=0; \,\,\,\ \rho^a\nabla_a(D\ln\Omega^2)=0~.
\label{2.08} 
\end{equation}
Here we used the notation $\{T_1,T_2\}\equiv T_1DT_2 -T_2DT_1$, $\{R_1,R_2\} \equiv -1/\kappa[D(T_1DT_2 - T_2DT_1)+(T_1DT_2 - T_2DT_1)D(\ln\Omega^2)]$ and the bracket in (\ref{2.07}) is a Lie one. For the detailed expression of the Lie bracket, see Eq. (\ref{a1.02}) of Appendix \ref{AppA}. It is interesting to note that the first and the second conditions of the above are equivalent. For instance, if we choose a coordinate system such that the Killing vector of the seed metric is given by $\chi^a = (1,0,0,0)$, then $DT$ will lead to the derivative with respect to time coordinate. In that case $DT$ will be regular near the horizon provided $T$ is made of regular functions in this region. So the vanishing of the above quantities only depends on the structure $\rho^a\nabla_a$. Therefore, the first one implies the second. Note that the first two were appeared earlier in Carlip's analysis \cite{Carlip:1999cy}. While the last one is new. Similarly, if $\ln\Omega^2$ is regular, then we will also have
\begin{equation}
\rho^a\nabla_a\ln\Omega^2 = 0~.
\label{2.08n1} 
\end{equation}
Also from the last condition of (\ref{2.08}) and (\ref{2.08n1}), it is possible to originate two more useful conditions, which are given by
\begin{equation}
\rho^a\nabla_a(D\ln\Omega) =0; \,\,\,\ \frac{1}{\Omega}\rho^a\nabla_a\Omega = 0~,
\label{2.08n2} 
\end{equation}
respectively. The subsequent analysis will be done using the diffeomorphism vectors, given by (\ref{2.05}) and (\ref{2.06}). To obtain the required expressions near the horizon the conditions in Eqs. (\ref{2.08}), (\ref{2.08n1}) and (\ref{2.08n2}) will be imposed several times. Remember that all the conditions are valid near the horizon. The robustness of them can be explicitly checked by considering a particular form of the metric. This we will do later.

\section{\label{Noether}Noether charge, bracket among the charges and the central term}
   Here using the diffeomorphisms, given in the earlier section, the form of the charge, the bracket among the charges and then the central term will be obtained in the near horizon limit. The discussion will be a general one. The facts we shall use are: the seed metric is stationary and both the conformal metric and the seed metric are solutions of the same gravitational theory. In the process of the analysis we shall always express the bar quantities in terms of the unbar ones. This will simplify the calculation a lot. A similar technique has been adopted earlier in \cite{Faraoni:2014lsa,Majhi:2014hpa} to determine several important quantities. The discussion will be presented in three subsections. In the first subsection, the charge and the bracket for the conformal background will be expressed in terms of the unbar quantities. Next using our diffeomorphisms the charge will be evaluated near the horizon. Final subsection will contain the analysis of the bracket and expression for the central term.

\subsection{Relations between quantities defined on seed metric and the conformally related metric}
   In this paper, I mainly concentrate on GR theory in four spacetime dimensions. We impose that both the static seed and conformally related matrices are solutions of GR. For example, Sultana-Dyer (SD) black hole is conformally related to the Schwarzschild metric. Also it can be shown that SD is a solution of GR with two noninteracting fluids \cite{Sultana:2005tp}. We will discuss later more on this. Since to obtain the horizon entropy in the context of Noether charge only the gravity part is necessary \cite{Wald:1993nt,Iyer:1995kg}, it is obvious that the form of the Noether current and hence the charge will be identical for both the matrices. Therefore the Noether charge for the conformal background is given by \cite{Paddy}
\begin{equation}
\bar{Q}[\bar{\xi}]=\frac{1}{2}\int\sqrt{\bar{\sigma}}d\bar{\Sigma}_{ab} \bar{J}^{ab}[\bar{\xi}]~,
\label{3.01}
\end{equation}
where the Noether potential $\bar{J}^{ab}[\bar{\xi}] = \bar{\nabla}^a\bar{\xi}^b-\bar{\nabla}^b\bar{\xi}^a$. The surface element of the horizon is $d\bar{\Sigma}_{ab} = -d^2x_{\perp}(\bar{N}_a\bar{M}_b - \bar{N}_b\bar{M}_a)$. $\bar{N}^a$ and $\bar{M}^a$ are the spacelike and timelike unit normals, respectively satisfying $\bar{g}_{ab}\bar{N}^a\bar{N}^b = +1$ and $\bar{g}_{ab}\bar{M}^a\bar{M}^b = -1$. $x_{\perp}$ denotes the transverse coordinates while $\bar{\sigma}$ is the determinant of the transverse metric. The integration has to be performed on the horizon. Remember that the above expression, in contrary to \cite{Wald:1993nt,Iyer:1995kg}, is valid for any diffeomorphism vector $\bar{\xi}^a$ and it is off-shell. A brief discussion of the derivation of the potential has been presented in Appendix \ref{Appnew}. In this case, a definition of the bracket among the charges can be given by $[\bar{Q}_1,\bar{Q}_2]\equiv\pounds_{\bar{\xi}_1}\bar{Q}[\bar{\xi}_2]-\pounds_{\bar{\xi}_2}\bar{Q}[\bar{\xi}_1]$. This was introduced earlier in one of my papers \cite{Majhi:2011ws}. It was shown that the bracket can be casted in terms of the Noether current $\bar{J}^a[\bar{\xi}]$ in the following form \cite{Majhi:2011ws}:
\begin{equation}
[\bar{Q}_1,\bar{Q}_2]: = \int\sqrt{\bar{\sigma}} d\bar{\Sigma}_{ab}\Big[\bar{\xi}^a_2\bar{J}^b_1 - 
\bar{\xi}^a_1\bar{J}^b_2]
\label{3.02}
\end{equation}
where we used the notation $\bar{J}^a_1 = \bar{J}^a[\bar{\xi}_1]$. Next we will express the above quantities in terms of those for the seed background. That will simplify the calculation. Before going into to the next step, let me mention some important features of the above expression for the bracket. First of all, it is off-shell and valid for any diffeomorphism vector. Also it is general enough to include any covariant theory of gravity. Secondly, substitution of current for GR in (\ref{3.02}) leads to that given by Carlip \cite{Carlip:1999cy}. Finally, one does not need to impose any ad-hoc condition, like $\delta_{\xi_1}\xi^a_2=0$ \cite{Carlip:1999cy} or boundary value, like Dirichlet or Newmann \cite{Silva:2002jq}. For details, please see the discussions below Eq. (9) of \cite{Majhi:2011ws}. The only information is needed to derived (\ref{3.02}) is that the current can be expressed as covariant derivative of the potential.

   To proceed in this direction, first use the fact that both the spacelike and timelike vectors obey the unit norm condition on both spacetimes; i.e. $g_{ab}N^aN^b = +1$; $g_{ab}M^aM^b = -1$ and $\bar{g}_{ab}\bar{N}^a\bar{N}^b = +1$, $\bar{g}_{ab}\bar{M}^a\bar{M}^b = -1$. Then the transformations between them must be
\begin{eqnarray}
&&\bar{N^a} = \Omega^{-1}N^a; \,\,\,\,\ \bar{N}_a=\Omega N_a;
\nonumber
\\
&&\bar{M^a} = \Omega^{-1}M^a; \,\,\,\,\ \bar{M}_a=\Omega M_a~.
\label{3.03} 
\end{eqnarray}
This leads the transformation relation between the two surface elements as $d\bar{\Sigma}_{ab} = \Omega^2 d\Sigma_{ab}$. Then using $\bar{\xi}_a = \Omega^2\xi_a$, we obtain
\begin{eqnarray}
\bar{J}_{ab}[\bar{\xi}] &=& \bar{\nabla}_a\bar{\xi}_b - \bar{\nabla}_b\bar{\xi}_a 
= \partial_a\bar{\xi}_b - \partial_b\bar{\xi}_a
= \Omega^2(\partial_a\xi_b - \partial_b\xi_a)+\xi_b\nabla_a\Omega^2 - \xi_a\nabla_b\Omega^2
\nonumber
\\
&=&\Omega^2 J_{ab}[\xi]-K_{ab}[\xi]~,
\label{3.04}
\end{eqnarray}
where $K_{ab}[\xi] = \xi_a\nabla_b\Omega^2 - \xi_b\nabla_a\Omega^2$. Therefore one can find
\begin{equation}
\bar{J}^{ab}[\bar{\xi}] = \bar{g}^{ai}\bar{g}^{bj}\bar{J}_{ij}[\bar{\xi}] 
= \Omega^{-2}J^{ab}[\xi] - \Omega^{-4}K^{ab}[\xi]~.
\label{3.05} 
\end{equation}
Finally, substituting all these in (\ref{3.01}) we express the charge in terms of the unbar entities:
\begin{equation}
\bar{Q}[\bar{\xi}]=\frac{1}{2}\int\sqrt{\sigma}~\Omega^2d\Sigma_{ab} \Big(J^{ab}[\xi] +
2\xi^b\nabla^a(\ln\Omega^2) \Big)~,
\label{3.06} 
\end{equation}
where $\sqrt{\bar{\sigma}}=\Omega^2\sqrt{\sigma}$ has been used.

  Let us now concentrate on the bracket (\ref{3.02}). For that we need to find the transformation relation among the currents in both frames. Remember that the current is related to the potential by $J^a = \nabla_b J^{ab}$. Therefore, using (\ref{3.05}) and $\bar{g} = (\Omega^2)^4 g$ we find
\begin{eqnarray}
\bar{J}^a &=& \bar{\nabla}_b\bar{J}^{ab} = \frac{1}{\sqrt{-\bar{g}}}\partial_b\Big(\sqrt{-\bar{g}} \bar{J}^{ab}\Big)
\nonumber
\\
&=& \frac{J^a}{\Omega^2} + \frac{2}{\Omega^3}J^{ab}\nabla_b\Omega - \frac{1}{\Omega^4}\nabla_b K^{ab}~.
\label{3.07}
\end{eqnarray}
Hence the bracket (\ref{3.02}) takes the following form:
\begin{equation}
[\bar{Q}_1,\bar{Q}_2]: = \int\sqrt{\sigma}\Omega^2 d\Sigma_{ab}\Big[(\xi^a_2J^b_1 +  \xi^a_2K_1^b) - (1\leftrightarrow 2)]~,
\label{3.08}
\end{equation}
where $K^b$ is given by
\begin{equation}
K^b = \frac{2}{\Omega}J^{bc}\nabla_c\Omega - \frac{1}{\Omega^2}\nabla_c K^{bc}~.
\label{3.09}
\end{equation} 
In the next two subsections we will calculate (\ref{3.06}) and (\ref{3.09}) in the near horizon limit.

\subsection{Evaluation of the charge near the horizon}
    To find the expression for charge (\ref{3.06}) in the near horizon limit we shall use some important relations which are evaluated earlier in the appendix B of \cite{Majhi:2011ws}. Since $\chi_a$ and $\rho_a$, as mentioned above, are defined with respect to the stationary metric $g_{ab}$, the covariant derivative of them can be expressed as
\begin{eqnarray}
&&\nabla_a\chi_b = \frac{\kappa}{\chi^2}(\chi_a\rho_b-\chi_b\rho_a)~;
\label{3.13}
\\
&&\nabla_a\rho_b = \frac{\kappa}{\chi^2}(\chi_a\chi_b-\rho_a\rho_b)~.
\label{3.14}
\end{eqnarray}
For details, see Eq. (B.10) and (B.26) of \cite{Majhi:2011ws}. Remember that the above quantities and the calculations from now are valid upto the leading order in $\chi^2$. Since we are only interested near the horizon, they can be neglected.
Also it has been shown in \cite{Majhi:2011ws} that the surface element for the stationary case is given by $d\Sigma_{ab}=d^2x_{\perp}\mu_{ab}$, where
\begin{equation}
\mu_{ab} = -\frac{|\chi|}{\rho\chi^2}(\chi_a\rho_b-\chi_b\rho_a)~.
\label{3.10}
\end{equation}
Therefore, the first term of (\ref{3.06}) leads to
\begin{equation}
\mu_{ab}J^{ab} = -\frac{2|\chi|}{\rho\chi^2}(\chi_a\rho_b-\chi_b\rho_a)\nabla^a\xi^b~.
\label{3.11}
\end{equation}
The near horizon expression of $\nabla_b\xi_b$ has been evaluated in Appendix \ref{AppB} (See Eq. (\ref{3.17})). Using this  the right hand side of (\ref{3.11}) takes the following form:
\begin{equation}
\mu_{ab}J^{ab} = \mu_{ab}(J^{ab})_0 - \frac{2|\chi|}{\rho\kappa}\Big[(DT)(D\ln\Omega^2)+TD^2(\ln\Omega^2)\Big]~,
\label{3.18}
\end{equation}
where we used the notation that $(\dots)_0$ represents the stationary expression; i.e. when the conformal factor $\Omega$ is unity.
Next concentrate on the other term of (\ref{3.06}). The relevant quantity, we have to look at is $\mu_{ab}\xi^b\nabla^a\ln\Omega^2$. This turns out to be  
\begin{eqnarray}
\mu_{ab}\xi^b\nabla^a\ln\Omega^2 &=& -\Big[\frac{|\chi|\rho^2}{\rho\chi^2}R\chi_a - \frac{|\chi|}{\rho}T\rho_a\Big]\nabla^a\ln\Omega^2
\nonumber
\\
&=& -\Big[\frac{|\chi|}{\rho\kappa}(DT)(D\ln\Omega^2) + \frac{T|\chi|}{\rho\kappa}(D\ln\Omega^2)^2 - \frac{|\chi|T}{\rho}\rho^a\nabla_a\ln\Omega^2\Big]~,
\label{3.19}
\end{eqnarray}
where in the last step the expression for $R$ from (\ref{2.06}) has been used.

   Now substituting (\ref{3.18}) and (\ref{3.19}) in (\ref{3.06}) and then using the condition (\ref{2.08n1}), we obtain
\begin{eqnarray}
\bar{Q}[\bar{\xi}]&=&\frac{1}{2}\int\sqrt{\sigma}~\Omega^2d\Sigma_{ab} (J^{ab}[\xi])_0
- \int\sqrt{\sigma}~\Omega^2d^2x_{\perp}\frac{|\chi|}{\rho\kappa}\Big[ 
2(DT)(D\ln\Omega^2)+TD^2(\ln\Omega^2)
\nonumber
\\
&+& T(D\ln\Omega^2)^2\Big]~.
\label{3.20}  
\end{eqnarray}
For the future purpose, using 
\begin{eqnarray}
&&D\ln\Omega^2 = \frac{2D\Omega}{\Omega}~;
\nonumber
\\
&&\nabla_a(D\ln\Omega^2)=\frac{2\nabla_a(D\Omega)}{\Omega}-\frac{2(D\Omega)\nabla_a\Omega}{\Omega^2}~;
\nonumber
\\
&&\nabla_a\nabla_b(D\ln\Omega^2)=\frac{2\nabla_a\nabla_b(D\Omega)}{\Omega}-\frac{2(\nabla_bD\Omega)
\nabla_a\Omega}{\Omega^2}-\frac{2(\nabla_aD\Omega)\nabla_b\Omega}{\Omega^2}+\frac{4D\Omega}{\Omega^3}
\nabla_a\Omega\nabla_b\Omega
\nonumber
\\
&&~~~~~~~~~~~~~~~~~~~~~~~ -\frac{2D\Omega}{\Omega^2}\nabla_a\nabla_b\Omega~;
\nonumber
\\
&& D^2\ln\Omega^2 = \frac{2\kappa}{\Omega}\rho^a\nabla_a\Omega+\frac{2}{\Omega}\chi^a\chi^b\nabla_a\nabla_b\Omega
-\frac{2(D\Omega)^2}{\Omega^2}~;
\label{3.20n1}
\end{eqnarray}
and the conditions (\ref{2.08}), (\ref{2.08n1}) and (\ref{2.08n2}) in (\ref{3.20}), we find
\begin{eqnarray}
\bar{Q}[\bar{\xi}]&=&\frac{1}{2}\int\sqrt{\sigma}~\Omega^2d\Sigma_{ab} (J^{ab}[\xi])_0
- \int\sqrt{\sigma}~\Omega^2d^2x_{\perp}\frac{1}{\kappa}\Big[ 
\frac{4(DT)(D\Omega)}{\Omega}+\frac{2T}{\Omega}\chi^a\chi^b\nabla_a\nabla_b\Omega
\nonumber
\\
&+& 2T\frac{(D\Omega)^2}{\Omega^2}\Big]~.
\label{Q}   
\end{eqnarray}
We have written the above in such a way that the first term gives the charge for the stationary background and others are due to the conformal factor. This way of representation will save our calculation since the first term has already been calculated in \cite{Majhi:2011ws}. We shall just borrow it at the end. This trick will be followed again and again through out the paper.

\subsection{Evaluation of the bracket near the horizon and the central term}
  Let us now find the near horizon form of (\ref{3.08}). Since the current is covariant derivative of the potential, this contains terms like $\nabla_a\nabla_b\xi_c$. Therefore we shall first look for the expression of such term. The form of $\nabla_a\xi_b$ is given in (\ref{3.17}). Using this we obtain
\begin{equation}
\nabla_c\nabla_a\xi_b = (\nabla_c\nabla_a\xi_b)_0 + P_{cab}~,
\label{3.21} 
\end{equation}
where 
\begin{equation}
P_{cab} = \nabla_c\Big[\frac{\chi_a\rho_b}{\kappa\rho^2}(DT)(D\ln\Omega^2) + 
\frac{T\chi^2}{\kappa\rho^2}\rho_b\nabla_a(D\ln\Omega^2)+\frac{T}{\rho^2}
(\chi_a\chi_b-\rho_a\rho_b)(D\ln\Omega^2)\Big]~, 
\label{3.22}
\end{equation}
and $(\nabla_c\nabla_a\xi_b)_0$ again is defined with respect to the stationary background.  This has been calculated earlier. It is given by Eq. (B.43) of \cite{Majhi:2011ws}. For our purpose, let me give the explicit expression of this quantity:
\begin{equation}
(\nabla_c\nabla_a\xi_b)_0 = \frac{2\kappa}{\chi^4}\chi_a\chi_b\chi_cDT -\frac{1}{\kappa\chi^4}\chi_a\rho_b\chi_cD^3T-\frac{1}{\chi^4}\chi_a\chi_b\chi_cD^2T~.
\label{3.21n1}
\end{equation} 
Now as the current is given by $J^a = \nabla_bJ^{ab} = \nabla_b(\nabla^a\xi^b - \nabla^b\xi^a)$, using (\ref{3.21}) we obtain
\begin{equation}
J^b = (J^b)_0 + P_{c}^{~~[bc]}~,
\label{3.22n1} 
\end{equation}
where $(J^b)_0 = (\nabla_c\nabla^b\xi^c)_0 - (\nabla_c\nabla^c\xi^b)_0$ is the current with respect to the stationary background and $P_c^{~~[bc]} = P_c^{~~bc} - P_c^{~~cb}$. 
These two quantities have been calculated in Appendix \ref{AppB}. They are given by Eq. (\ref{3.23n1}) and (\ref{3.25}), respectively.
Use of them leads to,
\begin{eqnarray}
d\Sigma_{ab}\xi_2^aJ^b_1 - (1\leftrightarrow 2) &=& d\Sigma_{ab}\Big(\xi_2^aJ^b_1- 
(1\leftrightarrow 2)\Big)_0 + d^2x_{\perp}\mu_{ab}\Big[(T_2\chi^a+R_2\rho^a)P_{1c}^{~~[bc]}
\nonumber
\\
&+&\frac{\chi^2}{\kappa\rho^2}T_2\rho^a(J_1^b)_0(D\ln\Omega^2)- (1\leftrightarrow 2)\Big]~.
\label{3.23}
\end{eqnarray}
Here again we followed our earlier logic. The first term on the right hand side was calculated earlier in \cite{Majhi:2011ws} and we shall substitute the expression when it will be needed.
The above can be shown to be reduced to the following form in the near horizon limit (For details see, Appendix \ref{AppB}):
\begin{eqnarray}
d\Sigma_{ab}\xi_2^aJ^b_1 - (1\leftrightarrow 2) &=& d\Sigma_{ab}\Big(\xi_2^aJ^b_1- 
(1\leftrightarrow 2)\Big)_0 + d^2x_{\perp}\Big[\Big\{\frac{T_2D^2T_1}{\kappa}(D\ln\Omega^2)
\nonumber
\\
&-& \frac{T_2\nabla_cT_1}{\kappa}\rho^2\nabla^c(D\ln\Omega^2)
- \frac{T_1DT_2}{\kappa^2}\chi^b\rho^c\nabla_b\nabla_c(D\ln\Omega^2)
\nonumber
\\
&-&\frac{2T_1DT_2}{\kappa}(D^2\ln\Omega^2)\Big\}
- (1\leftrightarrow 2)\Big]~.
\label{3.27}
\end{eqnarray}
Next consider the other term of (\ref{3.08}). This comes out to be
\begin{eqnarray}
&&d\Sigma_{ab}\xi_2^aK_1^b - (1\leftrightarrow 2) = d^2x_{\perp}\Big[\frac{4(D\Omega)}{\Omega\kappa}T_2D^2T_1
+ \frac{4\chi^2}{\Omega^2\kappa}T_2DT_1\nabla_c\Omega\nabla^c\Omega 
+ \frac{4\chi^2}{\Omega\kappa}T_2DT_1\Box\Omega 
\nonumber
\\
&&+ \frac{2}{\Omega\kappa}(\rho^a\rho^bT_2DT_1
 + \chi^a\chi^bT_1DT_2)\nabla_a\nabla_b\Omega -\frac{6(D\Omega)^2}{\Omega^2\kappa}T_1DT_2\Big]
- (1\leftrightarrow 2)~.
\label{3.36}
\end{eqnarray}
For detailed derivation of the above, see Appendix \ref{AppB}.
Substitution of (\ref{3.27}) and (\ref{3.36}) in (\ref{3.08}) leads to the following expression for the bracket among the charges:
\begin{eqnarray}
&&[\bar{Q}_1,\bar{Q}_2]: = \int\sqrt{\sigma}\Omega^2 d\Sigma_{ab}\Big(\xi_2^aJ_1^b - (1\leftrightarrow 2)\Big)_0 
+ \int\sqrt{\sigma}\Omega^2d^2x_{\perp}\Big[\Big\{\frac{T_2D^2T_1}{\kappa}(D\ln\Omega^2)
\nonumber
\\
&& - \frac{T_2\nabla_cT_1}{\kappa}
\rho^2\nabla^c(D\ln\Omega^2) - \frac{T_1DT_2}{\kappa^2}\chi^b\rho^c\nabla_b\nabla_c(D\ln\Omega^2) 
- \frac{2T_1DT_2}{\kappa}(D^2\ln\Omega^2)
\nonumber
\\
&& +\frac{4D\Omega}{\Omega\kappa}T_2D^2T_1 + \frac{4\chi^2}{\Omega^2\kappa}T_2DT_1\nabla_c\Omega\nabla^c\Omega
+\frac{4\chi^2}{\Omega\kappa}T_2DT_1\Box\Omega + \frac{2}{\Omega\kappa}(\rho^a\rho^bT_2DT_1+\chi^a\chi^bT_1DT_2)
\nabla_a\nabla_b\Omega
\nonumber
\\
&& - \frac{6(D\Omega)^2}{\Omega^2\kappa}T_1DT_2\Big\}-(1\leftrightarrow 2)\Big]~.
\label{3.37}
\end{eqnarray}
Finally, use of (\ref{3.20n1}) and the near horizon conditions (\ref{2.08}), (\ref{2.08n1}) and (\ref{2.08n2}) for the closer relation the above reduces to
\begin{eqnarray}
&&[\bar{Q}_1,\bar{Q}_2]: = \int\sqrt{\sigma}\Omega^2 d\Sigma_{ab}\Big(\xi_2^aJ_1^b - (1\leftrightarrow 2)\Big)_0 
+ \int\sqrt{\sigma}\Omega^2d^2x_{\perp}\Big[\Big\{\frac{6T_2D^2T_1}{\kappa}\frac{D\Omega}{\Omega}
\nonumber
\\
&& - \frac{T_2\nabla_cT_1}{\kappa}
\rho^2\Big(\frac{2\nabla^cD\Omega}{\Omega}-\frac{2D\Omega}{\Omega^2}\nabla^c\Omega\Big)
 - \frac{T_1DT_2}{\kappa^2}\chi^b\rho^c\Big(\frac{2\nabla_b\nabla_c(D\Omega)}{\Omega}
-\frac{2D\Omega}{\Omega^2}\nabla_b\nabla_c\Omega\Big) 
\nonumber
\\
&&-\frac{2T_1DT_2}{\Omega\kappa}\chi^a\chi^b\nabla_a\nabla_b\Omega
-\frac{2T_1DT_2}{\Omega^2\kappa}\frac{(D\Omega)^2}{\Omega^2} + \frac{4\chi^2}{\Omega^2\kappa}T_2DT_1\nabla_c\Omega\nabla^c\Omega
+\frac{4\chi^2}{\Omega\kappa}T_2DT_1\Box\Omega 
\nonumber
\\
&&+ \frac{2T_2DT_1}{\Omega\kappa}\rho^a\rho^b\nabla_a\nabla_b\Omega\Big\}-(1\leftrightarrow 2)\Big]~.
\label{QQ}
\end{eqnarray}

Now we compute the central term, which is defined by the relation
\begin{equation}
\bar{K}[\bar{\xi}_1,\bar{\xi}_2] =  [\bar{Q}_1,\bar{Q}_2]-\bar{Q}[\{\bar{\xi}_1,\bar{\xi}_2\}]~.
\label{3.38}
\end{equation}
Therefore, using (\ref{Q}) and (\ref{QQ}), we find the central term as
\begin{eqnarray}
\bar{K}[\bar{\xi}_1,\bar{\xi}_2] &=& -\frac{1}{32\pi G}\int\sqrt{\bar{\sigma}}P^{abcd}\mu_{ab}\mu_{cd}\frac{1}{\kappa}
\Big[DT_1D^2T_2 - DT_2D^2T_1\Big]
\nonumber
\\
&+&\int\sqrt{\bar{\sigma}}d^2x_{\perp}
\Big[\Big\{\frac{2T_2D^2T_1}{\kappa}\frac{D\Omega}{\Omega}
- \frac{T_2\nabla_cT_1}{\kappa}
\rho^2\Big(\frac{2\nabla^cD\Omega}{\Omega}-\frac{2D\Omega}{\Omega^2}\nabla^c\Omega\Big)
\nonumber
\\
&-& \frac{T_1DT_2}{\kappa^2}\chi^b\rho^c\Big(\frac{2\nabla_b\nabla_c(D\Omega)}{\Omega}
-\frac{2D\Omega}{\Omega^2}\nabla_b\nabla_c\Omega\Big) 
+ \frac{4\chi^2}{\Omega^2\kappa}T_2DT_1\nabla_c\Omega\nabla^c\Omega
\nonumber
\\
&+& \frac{4\chi^2}{\Omega\kappa}T_2DT_1\Box\Omega 
+ \frac{2T_2DT_1}{\Omega\kappa}\rho^a\rho^b\nabla_a\nabla_b\Omega\Big\}-(1\leftrightarrow 2)\Big]~.
\label{CT} 
\end{eqnarray}
In the above the values of the terms like $(\dots)_0$ have been substituted from \cite{Majhi:2011ws} and for GR $P^{abcd} = 1/2(g^{ac}g^{bd} - g^{ad}g^{bc})$.
In the next section the Fourier modes of the (\ref{Q}) and (\ref{CT}) will be found out using the decomposition of the function $T$.

\section{\label{Fourier}Fourier modes of charge and central term}
   To find the Fourier modes for the charge (\ref{Q}) and the central term (\ref{CT}), we shall follow the identical steps as adopted in \cite{Majhi:2011ws}. First consider the Fourier modes of the function $T$. Let us decompose it as
\begin{equation}
T = \displaystyle\sum_{m}A_mT_m 
\label{4.01}
\end{equation}
with $A^{*}_m = A_{-m}$. $T_m$ will be chosen in such a way that $T$ is real and the Fourier modes of the diffeomorphism vectors $\bar{\xi}^a_{m}$ satisfy 
\begin{equation}
i\{\bar{\xi}_m,\bar{\xi}_n\}^a = (m-n)\bar{\xi}_{m+n}^a~.
\label{4.02} 
\end{equation}
The left hand side of the above is basically the Fourier modes of the Lie bracket given in (\ref{2.07}). Substituting (\ref{4.01}) in (\ref{Q}) and using the decomposition $\bar{Q}[\bar{\xi}]=
\displaystyle\sum_mA_m\bar{Q}_m$, we obtain
\begin{eqnarray}
\bar{Q}_m &=& -\frac{1}{32\pi G}\int\sqrt{\bar{\sigma}}d^2x_{\perp}P^{abcd}\mu_{ab}\mu_{cd}\Big[2\kappa T_m 
-\frac{1}{\kappa}D^2T_m\Big]
\nonumber
\\
&-& \int\sqrt{\sigma}~\Omega^2d^2x_{\perp}\frac{1}{\kappa}\Big[ 
\frac{4(DT_m)(D\Omega)}{\Omega}+\frac{2T_m}{\Omega}\chi^a\chi^b\nabla_a\nabla_b\Omega
+ 2T_m\frac{(D\Omega)^2}{\Omega^2}\Big]~,
\label{Qm}
\end{eqnarray}
where $(\dots)_0$ was substituted from \cite{Majhi:2011ws}.
Similarly, using the modes for $T_1$ and $T_2$ and decomposing the certral term as $\bar{K}[\bar{\xi}_1,\bar{\xi}_2]=\displaystyle\sum_{m,n}C_{m,n}\bar{K}[\bar{\xi}_m,\bar{\xi}_n]$ in (\ref{CT}), we obtain
\begin{eqnarray}
\bar{K}[\bar{\xi}_m,\bar{\xi}_n] &=& -\frac{1}{32\pi G}\int\sqrt{\bar{\sigma}}P^{abcd}\mu_{ab}\mu_{cd}\frac{1}{\kappa}
\Big[DT_mD^2T_n - DT_nD^2T_m\Big]
\nonumber
\\
&+&\int\sqrt{\bar{\sigma}}d^2x_{\perp}
\Big[\Big\{\frac{2T_nD^2T_m}{\kappa}\frac{D\Omega}{\Omega}
- \frac{T_n\nabla_cT_m}{\kappa}
\rho^2\Big(\frac{2\nabla^cD\Omega}{\Omega}-\frac{2D\Omega}{\Omega^2}\nabla^c\Omega\Big)
\nonumber
\\
&-& \frac{T_mDT_n}{\kappa^2}\chi^b\rho^c\Big(\frac{2\nabla_b\nabla_c(D\Omega)}{\Omega}
-\frac{2D\Omega}{\Omega^2}\nabla_b\nabla_c\Omega\Big) 
+ \frac{4\chi^2}{\Omega^2\kappa}T_nDT_m\nabla_c\Omega\nabla^c\Omega
\nonumber
\\
&+& \frac{4\chi^2}{\Omega\kappa}T_nDT_m\Box\Omega 
+ \frac{2T_nDT_m}{\Omega\kappa}\rho^a\rho^b\nabla_a\nabla_b\Omega\Big\}-(m\leftrightarrow n)\Big]~.
\label{Km}  
\end{eqnarray}

To proceed further we need to choose the explicit expression for $T_m$. As explained above, it must be chosen such that they satisfy the algebra (\ref{4.02}). Remember that the seed metric is taken to be static and hence it is possible to choose a coordinate system in which the Killing vector takes the form $\chi^a=(1,0,0,0)$. Moreover, this is also the conformal Killing vector for the conformally related spacetime. Then the ansatz for $T_m$, in ($t,r,x_{\perp}$) coordinates, is
\begin{equation}
T_m = \frac{1}{\alpha}\textrm{exp}\Big[im(\alpha t+g(r)+p.x_{\perp})\Big]~.
\label{Tm} 
\end{equation}
Here $\alpha$ is a constant and the function $g(r)$ is regular near the horizon. It can be easily checked by using (\ref{2.07}) that the above satisfies (\ref{4.02}). Now since $T$ is a smooth function which is periodic and the Euclidean time coordinate has a periodicity $2\pi/\kappa$, the value of $\alpha$ must be $\alpha=\kappa$. Therefore we can substitute $\alpha$ by $\kappa$ in the above without any loss of generality. The same has been followed earlier in \cite{Silva:2002jq,Majhi:2011ws}, and recently in \cite{Majhi:2013lba}. Before concluding this section, I mention that (\ref{Qm}) and (\ref{Km}) are general enough to include any metric solution in GR which is conformally related to stationary metric, provided the conditions (\ref{2.08}), (\ref{2.08n1}) and (\ref{2.08n2}) are satisfied.

\section{\label{Sultana}Virasoro algebra and entropy of Sultana-Dyer black hole}
   In this section, we shall calculate the Fourier modes of the charge and the central term explicitly for Sultana-Dyer (SD) metric. The algebra will be shown to be identical to the standard Virasoro algebra. Then using Cardy formula the entropy associated to the conformal Killing horizon of the SD metric will be evaluated.

   Let me first introduce the Sultana-Dyer (SD) black hole briefly. It is a solution of general relativity (GR) with two sources. These are two noninteracting perfect fluids: one is timelike and the other one is null-like. The asymptotic form of the metric is Friedmann-Lemaitre-Robertson-Walker (FLRW). Moreover, the spacetime is conformal to the Schwarzschild black hole. For details and several features, see \cite{Sultana:2005tp}. The form of the line element is of the following form \cite{Sultana:2005tp}:
\begin{equation}
ds^2 = a^2(\eta)\Big[-d\eta^2+dr^2+r^2(d\theta^2+\sin^2\theta d\phi^2)+\frac{2m}{r}(d\eta+dr)^2\Big]~,
\label{SD1} 
\end{equation}
where $m$ is identified as the mass of the Schwarzschild black hole. $\eta$, $r$ are the time and the radial coordinates, respectively. The conformal factor is given by $a(\eta) = \eta^2$. Under the coordinate transformation $\eta = t + 2m \ln(r/2m - 1)$ the above takes the following form:
\begin{equation}
ds^2 =  a^2(t,r)\Big[-\Big(1-\frac{2m}{r}\Big)dt^2 + \frac{dr^2}{\Big(1-\frac{2m}{r}\Big)}+ r^2(d\theta^2
+\sin^2\theta d\phi^2)\Big]~,
\label{SDSC}
\end{equation}
in which case the conformal factor is \cite{Faraoni:2013aba},
\begin{equation}
a(t,r) = \Big(t+2m\ln\Big|\frac{r}{2m}-1\Big|\Big)^2~.
\label{a}
\end{equation}
This tells that the SD metric is conformal to the usual Schwarzschild metric. It has a conformal Killing vector $\bar{\chi}^a = (1,0,0,0)$, which is the Killing vector for the Schwarzschild spacetime. The co-vector is given by $\bar{\chi}_a = a^2(-F,0,0,0)$ where $F=(1-2m/r)$. Therefore, the horizon of the SD metric, denoted by $\bar{\chi}^2=0$, is $r=2m$. Since, the conformal factor is time dependent, the spacetime is an evolving one.

     Before proceeding further let me point out an important feature of the SD metric. As I mentioned earlier, it is a solution of GR with two sources: one is timelike dust and other is null fluid. The total energy momentum tensor for these are $T_{ab} = \mu u^au^b + \tau k^ak^b$, where the first term is for timelike dust with energy density $\mu$ and zero pressure while last term is for null source. Here $u^a$ is the four velocity which is timelike and on the other hand $k^a$ is null like. The time and radial components of these vectors, which are only non-zero, are given by \cite{Sultana:2005tp},
\begin{eqnarray}
&&u^0 = \frac{r^2+m(2r-\eta)}{r\eta^2\sqrt{r^2+2m(r-\eta)}}~; \,\,\,\ u^1 = \frac{m(\eta-2r)}{r\eta^2\sqrt{r^2+2m(r-\eta)}}~;
\nonumber
\\
&& k^0 = \frac{\sqrt{2m(r-\eta)+r^2}}{r\eta^2}~; \,\,\,\ k^1 = - \frac{\sqrt{2m(r-\eta)+r^2}}{r\eta^2}~.
\label{uk}
\end{eqnarray}
The energy density of the dust turns out to be \cite{Sultana:2005tp},
\begin{equation}
\mu = \frac{12(r^2+2m(r-\eta))}{r^2\eta^6}~,
\label{mu}
\end{equation}
which is positive in the region
\begin{equation}
\eta<\frac{r(r+2m)}{2m}~.
\label{eta}
\end{equation}
Now note that on the horizon $r=2m$ both $u^0$ and $k^0$ are positive while both $u^1$ and $k^1$ are negative for (\ref{eta}). That means if the energy condition is satisfied, an observer who is outside the horizon, will see that the dust and the null fluid flow radially into the black hole. On the other hand, for late times; i.e. for $\eta>\frac{r(r+2m)}{2m}$ the sources become unphysical and the dust becomes superluminal. When $r=2m$, the energy condition is obeyed for $\eta<4m$, otherwise the particles near the horizon will be superluminal. In spite of such unphysical feature of SD solution, it is still interesting as the global structure is similar to that of a cosmological black hole to represent a more realistic situation. Moreover, the metric is simple enough to explore the thermodynamics and other aspects to gather more information about our universe. One might think that this unphysical superluminal feature exists as it is far from the realistic one. However, one can expect that an exact solution will be free of this problem. In absence of such solutions, here I shall adopt the SD metric model with the expectation that it will represent the similar situation of the real world.

   Being time dependent, the usual stationary way of defining thermodynamical quantities does not work here. To obtain a full thermodynamics of such spacetimes, one needs to correctly obtain each entities from the first principle. For instance, in \cite{Saida:2007ru,Majhi:2014hpa} the temperature was determined by calculating the radiation spectrum from the horizon. But in literature no derivation of entropy exists. People usually assume the usual area law without any proper justification \cite{Faraoni:2007gq,Majhi:2014hpa}. One might argue that since SD is a solution of GR, its entropy will be given by area law. But this has been established rigorously for the stationary cases. So without any direct calculation, the validity of this argument can be falsified. In the below, I will obtain the entropy of SD spacetime in the context of Virasoro algebra and Cardy formula by using the modes of charge and central term, obtained in the previous section. This will provide the first direct calculation of the entropy for the present case.    

    Note that the conformal factor $\Omega=a$, given in (\ref{a}), diverges at $r=2m$. Therefore to use (\ref{Qm}) and (\ref{Km}), which are based on the conditions (\ref{2.08}), (\ref{2.08n1}) and (\ref{2.08n2}), it is necessary to check if the conditions are still hold near the horizon. Using first equation of (\ref{2.02}) we obtain the form of $\rho^a$ as
\begin{equation}
\rho_a = (0,\frac{F'}{2\kappa},0,0); \,\,\,\ \rho^a=(0,\frac{FF'}{2\kappa},0,0)~.
\label{5.01} 
\end{equation}
Since $\chi^a=(1,0,0,0)$, we find $\rho^a\nabla_aDT = (FF'/2\kappa)\partial_r\partial_t T$ which vanishes at horizon as $F(r=2m)=0$ and $\partial_r\partial_t T$ is finite. Similarly, the first condition of (\ref{2.08}) is also satisfied in the near horizon limit. Let us now calculate $\rho^a\nabla_a(D\ln\Omega^2)$. This, for the present case, turns out to be $\rho^a\nabla_a(D\ln\Omega^2) = -4mF'/\kappa ra$. Since $a$ diverges near the horizon, it vanishes in this limit. In the identical way, it is also possible to show that the other conditions; i.e. (\ref{2.08n1}) and (\ref{2.08n2}) are also valid. Therefore for SD metric, we can use the expressions for the Fourier modes of the charge (\ref{Qm}) and the central term (\ref{Km}). To evaluate (\ref{Qm}) and (\ref{Km}), we concentrate on the terms other than the first term. Using the expression for the conformal factor (\ref{a}) and using the fact that it diverges near the horizon, it is possible to show that these vanish in the near horizon limit. Hence the modes for the charge and the central term reduces to the following forms:
\begin{eqnarray}
&&\bar{Q}_m = -\frac{1}{32\pi G}\int\sqrt{\bar{\sigma}}d^2x_{\perp}P^{abcd}\mu_{ab}\mu_{cd}\Big[2\kappa T_m 
-\frac{1}{\kappa}D^2T_m\Big]~;
\label{5.02}
\\
&&\bar{K}[\bar{\xi}_m,\bar{\xi}_n] = -\frac{1}{32\pi G}\int\sqrt{\bar{\sigma}}d^2x_{\perp}P^{abcd}\mu_{ab}\mu_{cd}\frac{1}{\kappa}
\Big[DT_mD^2T_n - DT_nD^2T_m\Big]~.
\label{5.03}
\end{eqnarray}
Note that these are identical to the stationary case except the measure of the integration has been modified by the conformal factor.
Next substituting (\ref{Tm}) in the above and then integrating over the transverse coordinates we obtain
\begin{eqnarray}
&&\bar{Q}_m = \frac{\bar{A}}{8\pi G}\delta_{m,0}~;
\label{QmSD}
\\
&& \bar{K}[\bar{\xi}_m,\bar{\xi}_n] =-im^3\frac{\bar{A}}{8\pi G}\delta_{m+n,0}~,
\label{KmSD}
\end{eqnarray}
where $\bar{A} = -1/2\int\sqrt{\sigma}a^2d^2x_{\perp}P^{abcd}\mu_{ab}\mu_{cd} = 16\pi a^2m^2$ is the horizon area. 
It may be noted that the above two expressions satisfy the following algebra
\begin{equation}
i[\bar{Q}_m,\bar{Q}_n] = (m-n)\bar{Q}_{m+n}+m^3\frac{\bar{A}}{8\pi G}\delta_{m+n,0}~.
\label{5.04} 
\end{equation}
This is the standard form of the Virasoro algebra in which the zero mode eigenvalue ($\bar{Q}_0$) and the central charge ($\bar{C}$) are identified as
\begin{equation}
\frac{\bar{C}}{12} =\frac{\bar{A}}{8\pi G}; \,\,\,\ \bar{Q}_0 =  
\frac{\bar{A}}{8\pi G}~.
\label{CQ} 
\end{equation}
Now the standard form of the Cardy formula for entropy ($\bar{S}$) is given by \cite{Cardy:1986ie,Carlip:1998qw}
\begin{equation}
\bar{S} = 2\pi \sqrt{\frac{\bar{C}\Delta}{6}}; \,\,\,\,\ \Delta\equiv \bar{Q}_0-\frac{\bar{C}}{24}~,
\label{Cardy}  
\end{equation}
which leads to the following form of the entropy as
\begin{equation}
\bar{S}=\frac{\bar{A}}{4G}~.
\label{S} 
\end{equation}
This is the standard Bekenstein-Hawking area law for entropy. In earlier works \cite{Faraoni:2007gq,Majhi:2014hpa} on thermodynamics of SD metric, such form of entropy has been taken without any proper justification. Whereas the present calculation gives a direct estimation of the entropy of the conformal Killing horizon of SD metric.

\section{\label{Conclusions}Conclusions and discussions}
      The idea of calculating horizon entropy from the Noether charge in the context of Virasoro algebra and Cardy formula by using the diffeomorphisms for the invariance of the horizon structure has been given by Carlip \cite{Carlip:1999cy}. This has been followed by several people. But almost all the attempts were confined for the stationary spacetimes. In this paper, the black holes were chosen which are conformal to the stationary solutions in GR. Since the conformal factor is a function of time, the metric is no longer stationary. Till now no attempt has been done for the time dependent case in the present context. Moreover, it is not obvious if the area law for entropy in the case of GR remains invariant under conformal transformation. Here we have addressed these issues.

    Using the fact that the Killing vector of the seed stationary metric acts as the conformal Killing vector for its conformal partner and the horizon is defined by the vanishing of the norm of it, we determined a class of diffeomorphism vectors which leave the horizon structure invariant. The Fourier modes of the Noether charge and the central term were obtained form these set of vectors. Upto this our discussion was general on the basis of the following facts. (i) Both matrices are solutions of GR and one is conformal to the other with the seed metric is stationary. (ii) The conformal factor is a general function of all the coordinates. (iii) There are certain conditions, given by (\ref{2.08}), (\ref{2.08n1}) and (\ref{2.08n2}), must be satisfied by the conformal factor and the function $T$. Therefore, these results can be used for any conformal spacetime which obey the above facts. In this paper, we have considered the Sultana-Dyer (SD) metric which is conformal to the Schwarzschild metric and a solution of GR with two noninteracting fluids as sources. The conformal factor being time dependent, it is an evolving solution. Calculating the modes for the charge and the central term for SD case, we showed that these satisfy the Virasoro like algebra. Then use of Cardy formula led to the entropy.

   Now let us discuss what we have achieved in this paper. First of all this gives a way to handle the spacetimes which are conformal to the stationary ones for calculating the entropy by Cardy formula. Secondly, since the conformal factor is time dependent, we discussed the evolving situation. This is important because our universe is not static and so it is necessary to know how to study the thermodynamics of the realistic situations to gain better understanding and information of the universe. Finally, there does not exist any direct calculation of the entropy of the SD  metric in literature. People usually considered the area law without any concrete justification \cite{Faraoni:2007gq,Majhi:2014hpa}. The present calculation gave the first confirmation of this assumption. Hence I believe that this paper is an important footstep towards the current issues.

   Finally, it is worth to mention that in principle the thermodynamics of the dynamical solutions must be handled by a most sophisticated way. If the evaluation is very slow, then one can use the ``adiabatic approximation'' technique. On the other hand if the solution has violent dynamics, then it is necessary to develop the ``non-equilibrium technique'' to study the thermodynamics. Unfortunately, till now there is no proper frame work of such theory and hence it is hard to extract any definite answer in these cases. More definitely, it is almost impossible to treat the thermodynamics of any dynamical solution if the evaluation is not smooth enough. Here I have avoided such complexities. This has been possible as the solutions are related to our known stationary solutions by time dependent conformal factor. Of course, it is hoped that the concepts of thermodynamics of the evolving solutions will be more clear when one can develop the proper theoretical tool to handle these cases.

\vskip 9mm
\noindent 
{\bf Acknowledgements}\\
The research of the author is supported by a Lady Davis Fellowship at Hebrew University, by the I-CORE Program of the Planning and Budgeting Committee and the Israel Science Foundation (grant No. 1937/12), as well as by the Israel Science Foundation personal grant No. 24/12.

\vskip 6mm
\section*{Appendices}
\appendix
\section{\label{AppA}Calculation of Lie bracket}
\renewcommand{\theequation}{A.\arabic{equation}}
\setcounter{equation}{0} 
The Lie bracket $\pounds_{\bar{\xi}_1}\bar{\xi}^a_2=\{\bar{\xi}_1,\bar{\xi}_2\}^a$, using (\ref{2.05}), 
turns out to be 
\begin{eqnarray}
\{\bar{\xi}_1,\bar{\xi}_2\}^a &=& \xi_1^b\partial_b\xi_2^a-\xi_2^b\partial_b\xi_1^a
\nonumber
\\
&=& \Big(\chi^aT_1DT_2 + T_1R_2D\rho^a+\rho^aT_1DR_2+R_1T_2\rho^b\nabla_b\chi^a + \chi^a\rho^bR_1\nabla_bT_2
\nonumber
\\
&+&\rho^a\rho^bR_1\nabla_bR_2\Big) - (1\leftrightarrow 2)~.
\label{a1.01}
\end{eqnarray}
Then we proceed further by substituting (\ref{3.13}), (\ref{3.14}) and (\ref{2.06}) to reach:
\begin{eqnarray}
\{\bar{\xi}_1,\bar{\xi}_2\}^a &=& (T_1DT_2 - T_2DT_1)\chi^a - \frac{1}{\kappa}\Big[\Big(D(T_1DT_2 - T_2DT_1)\Big)
+ (T_1DT_2 - T_2DT_1)(D\ln\Omega^2)\Big]\rho^a
\nonumber
\\
&+& \Big[\Big\{R_1\chi^a\rho^b\nabla_bT_2 - \frac{R_1}{\kappa}\rho^a\rho^b\nabla_b(DT_2) - \frac{R_1}{\kappa}\rho^a\rho^b
(\nabla_bT_2)(D\ln\Omega^2)
\nonumber
\\
&-& \frac{R_1T_2}{\kappa}\rho^a\rho^b\nabla_b(D\ln\Omega^2)\Big\}-(1\leftrightarrow 2)\Big]~.
\label{a1.02}
\end{eqnarray}
This has been used to obtain the near horizon conditions, given in (\ref{2.08}).

\section{\label{Appnew} Noether current and potential}
\renewcommand{\theequation}{B.\arabic{equation}}
\setcounter{equation}{0}  
   The explicit variation of the covariant Lagrangian $L(\bar{g}_{ab}, \bar{R}_{abcd})$ under the metric variation is given by
\begin{equation}
\delta(L\sqrt{-\bar{g}})= \sqrt{-\bar{g}}\Big[\bar{E}_{ab}\delta\bar{g}^{ab}+\bar{\nabla}_a\delta v^a\Big]
\label{new1}
\end{equation}
where the first term leads to the equation of motion and the last term is the surface contribution. Now if this variation is due to a diffeomorphism $x^a\rightarrow x^a+\bar{\xi}^a$, then the right hand side of the above can be expressed as $\sqrt{-\bar{g}}\Big[\bar{\nabla}_a(-2\bar{E}^a_b\bar{\xi}^b+\delta v^a)\Big]$. On the other hand, since the Lagrangian is a scalar, we have $\delta(L\sqrt{-\bar{g}}) = \sqrt{-\bar{g}}\bar{\nabla}_a(L\bar{\xi}^a)$. Equating both sides one finds the conservation condition $\bar{\nabla}_a\bar{J}^a=0$, where the Noether current is given by
\begin{equation}
\bar{J}^a = L\bar{\xi}^a+2\bar{E}^{ab}\bar{\xi}_b-\delta v^a~.
\label{new2}
\end{equation}
In the above the derivation is off-shell; i.e. no use of equation of motion has been done. Also note that the current is conserved automatically where no information of equation of motion is needed. Therefore the whole analysis is off-shell and is valid for any diffeomorphism vector $\bar{\xi}^a$. Since $\bar{J}^a$ is conserved, one can define an anti-symmetric tensor $\bar{J}^{ab}$ such that $\bar{J}^{a}=\bar{\nabla}_b\bar{J}^{ab}$. $\bar{J}^{ab}$ is called the Noether potential.
In the case of the covariant Lagrangian, these are coming out to be \cite{Paddy}
\begin{eqnarray}
\bar{J}^{a}&=&2\bar{P}^{abcd}\bar{\nabla} _{b}\bar{\nabla} _{c}\bar{\xi} _{d}-2\bar{\nabla} _{b}
\left(\bar{P}^{adbc}+\bar{P}^{acbd}\right)\bar{\nabla} _{c}\bar{\xi} _{d}-4\bar{\xi} _{d}\bar{\nabla} _{b}\bar{\nabla} _{c}\bar{P}^{abcd}~;
\label{vir08}
\\
\bar{J}^{ab}&=&2\bar{P}^{abcd}\bar{\nabla} _{c}\bar{\xi} _{d}-4\left(\bar{\nabla} _{c}\bar{P}^{abcd} \right)\bar{\xi} _{d}~,
\label{vir09}
\end{eqnarray}
where $\bar{P}^{abcd}=\partial L/\partial \bar{R}_{abcd}$.
For GR theory, substitution of each term in the above leads to $\bar{J}^a = \bar{\nabla}_b\bar{J}^{ab}$ where $\bar{J}^{ab}=\bar{\nabla}^a\bar{\xi}^b - \bar{\nabla}^b\bar{\xi}^a$. For the details, see the discussion given in page $394$ of the book \cite{Paddy}. Remember that the expressions for current and potential are general and can be used for any diffeomorphism vector.

\section{\label{AppB} Near horizon expressions of some required quantities}
\renewcommand{\theequation}{C.\arabic{equation}}
\setcounter{equation}{0}  
{\bf Evaluation of $\nabla_a\xi_b$ near the horizon:}\\
Using $\xi_a=T\chi_a+R\rho_a$ we obtain,
\begin{equation}
\nabla_a\xi_b = \nabla_a(T\chi_b+R\rho_b) = \chi_b\nabla_aT + T\nabla_a\chi_b+\rho_b\nabla_aR+R\nabla_a\rho_b~.
\label{3.12}
\end{equation}
Next use of (\ref{2.06}) leads to
\begin{equation}
\nabla_a R = (\nabla_a R)_0 + \frac{1}{\kappa\rho^2}\chi_a(DT)(D\ln\Omega^2) + \frac{T\chi^2}{\kappa\rho^2}(\nabla_a\ln\Omega^2)~,
\label{3.15}
\end{equation}
where $(\nabla_a R)_0$ was calculated earlier in \cite{Majhi:2011ws}. It is given by
\begin{equation}
(\nabla_a R)_0 = -\frac{1}{\kappa\chi^2}\chi_a(D^2T)~.
\label{3.16}
\end{equation}
To obtain (\ref{3.15}), the identity $\nabla_a T = (1/\chi^2)\chi_aDT$ was used. The proof of it is given in \cite{Majhi:2011ws} (see Eq. (B.13) of this Ref.).
Substitution of all these in (\ref{3.12}) yields,
\begin{equation}
\nabla_a\xi_b = (\nabla_a\xi_b)_0 + \frac{\chi_a\rho_b}{\kappa\rho^2}(DT)(D\ln\Omega^2)+ \frac{T\chi^2}{\kappa\rho^2}\rho_b\nabla_a(D\ln\Omega^2) + \frac{T}{\rho^2}(\chi_a\chi_b - \rho_a\rho_b)(D\ln\Omega^2)~,
\label{3.17}
\end{equation}
where $(\nabla_a\xi_b)_0$ is given by Eq. (B.37) of \cite{Majhi:2011ws}. For completeness, the expression is given below:
\begin{equation}
(\nabla_a\xi_b)_0 = \frac{\chi_a\chi_b}{\chi^2}DT + \frac{\kappa}{\chi^2}(\chi_a\rho_b-\chi_b\rho_a)T -\frac{1}{\kappa\chi^2}\chi_a\rho_bD^2T -\frac{1}{\chi^2}DT(\chi_a\chi_b-\rho_a\rho_b)
\label{3.17n1}
\end{equation}

\vskip 9mm
\noindent
{\bf Expression for $P_c^{~~[bc]}$:}\\
Use of (\ref{3.22}) and then an extensive calculation yields
\begin{eqnarray}
P_{c}^{~~[bc]} &=& \nabla_c\Big[\frac{1}{\kappa\rho^2}(\chi^b\rho^c - \chi^c\rho^b)(DT)(D\ln\Omega^2)
\nonumber
\\ 
&+& \frac{T\chi^2}{\kappa\rho^2}\Big\{\rho^c\nabla^b(D\ln\Omega^2)-\rho^b\nabla^c(D\ln\Omega^2)\Big\}\Big]
\nonumber
\\
&=& \frac{1}{\kappa\rho^2}\chi^b(\rho^c\nabla_cDT)(D\ln\Omega^2) + \frac{DT}{\kappa\rho^2}
\chi^b\rho^c\nabla_c(D\ln\Omega^2)-\frac{D^2T}{\kappa\rho^2}\rho^b(D\ln\Omega^2)
\nonumber
\\
&-&\frac{DT}{\kappa\rho^2}\rho^b(D^2\ln\Omega^2)+\frac{\chi^2}{\kappa\rho^2}(\rho^c\nabla_cT)\nabla^b(D\ln\Omega^2)
-\frac{\chi^2}{\kappa\rho^2}\rho^b(\nabla_cT)\nabla^c(D\ln\Omega^2)
\nonumber
\\
&-& 2T\nabla^b(D\ln\Omega^2)+\frac{T\chi^2}{\kappa\rho^2}\rho^c\nabla_c\nabla^b(D\ln\Omega^2)-\frac{T}{\rho^2}\chi^b
(D^2\ln\Omega^2)+\frac{T}{\rho^2}\rho^b\rho^c\nabla_c(D\ln\Omega^2)
\nonumber
\\
&-&\frac{T\chi^2}{\kappa\rho^2}\rho^b\Box(D\ln\Omega^2)~.
\label{3.25}
\end{eqnarray}
In the above, (\ref{3.13}) and (\ref{3.14}) was used in the intermediate steps.

\vskip 9mm
\noindent
{\bf Derivation of Eq. (\ref{3.27}):}\\
Since $(J^b)_0 =(\nabla_c\nabla^b\xi^c)_0 - (\nabla_c\nabla^c\xi^b)_0$ and $(\nabla_d\nabla_a\xi_b)_0$ 
is given by (\ref{3.21n1}), we find
\begin{equation}
(J^b)_0 = \Big(\frac{1}{\kappa\chi^2}D^3T - \frac{2\kappa}{\chi^2} DT\Big)\rho^b~.
\label{3.23n1}
\end{equation}
Then use of (\ref{3.10}) leads to the vanishing of $\mu_{ab}\rho^a(J_1^b)_0$ term.
Next using (\ref{3.10}) we find
\begin{eqnarray}
\mu_{ab}(T_2\chi^a+R_2\rho^a)P_{1c}^{~~[bc]} - (1\leftrightarrow 2) 
= -\frac{|\chi|}{\rho}(T_2\rho_b+R_2\chi_b)P_{1c}^{~~[bc]} - (1\leftrightarrow 2)~.
\label{3.24} 
\end{eqnarray}
where in the final step $\rho^2/\chi^2 = -1 + \mathcal{O}(\chi^2)$ has been used. Substitution of (\ref{3.25}) and (\ref{2.06}) in (\ref{3.24}) and imposition of the conditions 
(\ref{2.08}), (\ref{2.08n1}) and (\ref{2.08n2}) yields
\begin{eqnarray}
&&(T_2\rho_b+R_2\chi_b)P_{1c}^{~~[bc]} - (1\leftrightarrow 2) = \Big[- \frac{T_2D^2T_1}{\kappa}(D\ln\Omega^2)
+ \frac{T_2\nabla_cT_1}{\kappa}\rho^2\nabla^c(D\ln\Omega^2)
\nonumber
\\
&&+ \frac{T_1DT_2}{\kappa^2}\chi^b\rho^c\nabla_b\nabla_c(D\ln\Omega^2)+\frac{2T_1DT_2}{\kappa}(D^2\ln\Omega^2)\Big]
- (1\leftrightarrow 2)~.
\label{3.26}
\end{eqnarray}
Using the above in (\ref{3.23}), we obtain (\ref{3.27}).

\vskip 9mm
\noindent
{\bf Derivation of Eq. (\ref{3.36}):}\\
Using the expression for $K^a$ (see Eq. (\ref{3.09})) and $\Sigma_{ab}=d^2x_{\perp}\mu_{ab}$ we find 
\begin{eqnarray}
&&d\Sigma_{ab}\xi_2^aK_1^b - (1\leftrightarrow 2) = d^2x_{\perp}\mu_{ab}\xi_2^a 
\Big[-\frac{4}{\Omega}(\nabla^c\xi^b_1)(\nabla_c\Omega) - \frac{2}{\Omega^2}\xi_1^b
(\nabla_c\Omega)(\nabla^c\Omega)
\nonumber
\\
&& - \frac{2}{\Omega}\xi_1^b\Box\Omega
+ \frac{2}{\Omega}(\nabla_c\xi_1^c)(\nabla^b\Omega)+\frac{2}{\Omega^2}\xi_1^c
(\nabla_c\Omega)(\nabla^b\Omega)+\frac{2}{\Omega}\xi_1^c\nabla_c\nabla^b\Omega 
\nonumber
\\
&& + \frac{2}{\Omega}(\nabla^b\xi^c)(\nabla_c\Omega)\Big] - (1\leftrightarrow 2)~.
\label{3.28}
\end{eqnarray}
Below I give the final expressions of each of the above terms near the horizon:
\begin{eqnarray}
&&\mu_{ab}\xi_2^a(\nabla^c\xi_1^b)\nabla_c\Omega - (1\leftrightarrow 2) = 
-\frac{1}{\kappa}\Big[T_2D^2T_1 - (1\leftrightarrow 2)\Big](D\Omega)~;
\label{3.29}
\\
&& \mu_{ab}\xi_2^a\xi_1^b(\nabla_c\Omega)(\nabla^c\Omega) - (1\leftrightarrow 2) =  
-\frac{2\chi^2}{\kappa}\Big[T_2DT_1-(1\leftrightarrow 2)\Big](\nabla_c\Omega)(\nabla^c\Omega)~;
\label{3.30}
\\
&& \mu_{ab}\xi_2^a\xi_1^b\Box\Omega = -\frac{2\chi^2}{\kappa}\Big[T_2DT_1-(1\leftrightarrow 2)\Big]\Box\Omega~;
\label{3.31}
\\
&& \mu_{ab}\xi_2^a(\nabla_c\xi_1^c)(\nabla^b\Omega) = -\Big[\frac{T_1DT_2}{\kappa}
-  (1\leftrightarrow 2)\Big] (D\ln\Omega^2)(D\Omega)~;
\label{3.32}
\\
&& \mu_{ab}\xi_2^a\xi_1^c(\nabla_c\Omega)(\nabla^b\Omega) = 
\frac{1}{\kappa}\Big[T_1DT_2 -(1\leftrightarrow 2)\Big](D\Omega)^2~;
\label{3.33}
\\
&& \mu_{ab}\xi_2^a\xi_1^c\nabla_c\nabla^b\Omega = 
\frac{1}{\kappa}\Big[\Big(\rho^b\rho^cT_2DT_1+\chi^b\chi^cT_1DT_2\Big) - (1\leftrightarrow 2)\Big]\nabla_b\nabla_c\Omega~;
\label{3.34}
\\
&& \mu_{ab}\xi_2^a(\nabla^b\xi_1^c)(\nabla_c\Omega) = \frac{1}{\kappa}\Big[T_2DT_1 - (1\leftrightarrow 2)\Big]
(D\ln\Omega^2)(D\Omega)~.
\label{3.35}
\end{eqnarray}
Substitution of them in (\ref{3.28}) yields (\ref{3.36}).

\end{document}